\documentclass[onecolumn,prd,aps,preprintnumbers,floatfix,showpacs,preprint,tightenlines,nofootinbib]{revtex4}
\usepackage{epsfig}
\usepackage{graphics}
\usepackage{latexsym}
\usepackage{amsmath}
\usepackage{amssymb}
\usepackage{rotating}

\begin{document}
\preprint{JLAB-THY-08-840, ADP-08-07-T667, ANL-PHY-12093-TH-2008}

\title{Strange magnetic form factor of the proton at $Q^2 = 0.23$ GeV$^2$\footnote
{Notice: Authored by Jefferson Science Associates, LLC
under U.S. DOE Contract No. DE-AC05-06OR23177. The U.S. Government
retains a non-exclusive, paid-up, irrevocable, world-wide license to
publish or reproduce this manuscript for U.S. Government purposes.}}

\author{P. Wang$^{a}$}
\author{D. B. Leinweber$^b$}
\author{A. W. Thomas$^{a,c}$}
\author{R. D. Young$^d$}

\affiliation{ $^a$Jefferson Laboratory, 12000 Jefferson Ave.,
Newport News, VA 23606 USA}

\affiliation{ $^b$Special Research Center for the Subatomic
Structure of Matter (CSSM) and Department of Physics, University of
Adelaide 5005, Australia}

\affiliation{$^c$Department of Physics, College of William and Mary, Williamsburg VA 23187 USA}

\affiliation{ $^d$Physics Division, Argonne National Laboratory,
Argonne, IL 60439, USA}

\begin{abstract}
We determine the $u$ and $d$ quark contributions to the proton
magnetic form factor at finite momentum transfer by applying chiral
corrections to quenched lattice data. Heavy baryon chiral
perturbation theory is applied at next to leading order in the
quenched, and full QCD cases for the valence sector using finite
range regularization. Under the assumption of charge symmetry these
values can be combined with the experimental values of the proton
and neutron magnetic form factors to deduce a relatively accurate
value for the strange magnetic form factor at $Q^2=0.23$ GeV$^2$,
namely $G_M^s=-0.034 \pm 0.021$ $\mu_N$.
\end{abstract}
\pacs{13.40.-f; 14.20.-c 12.39.Fe; 11.10.Gh}
%
%
\maketitle
%

Strange quark contributions to the properties of the nucleon have
attracted a lot of interest since the originally puzzling EMC
results concerning the proton spin~\cite{EMC}. While that motivation
has faded~\cite{Myhrer:2007cf,Thomas:2008bd}, it remains a central
issue in QCD, especially with respect to lattice QCD, where such
terms necessarily involve so-called ``disconnected graphs'', i.e.,
quark loops which are connected only by gluon lines to the valence
quarks. Despite enormous effort~\cite{Lewis:2002ix}, the {\it
direct} lattice calculations of these contributions have so far been
unable to produce a result which differs statistically from zero. On
the other hand, by using the constraints of charge symmetry, which
is expected to be accurate at the 1\% level or
better~\cite{Miller:1997ya,Miller:1990iz}, one can write relations
(c.f. Eqs.~(\ref{magp})  and (\ref{magn}), below) for the
disconnected contributions to physical form
factors~\cite{Leinweber:1999nf} in terms of valence quantities,
which {\it can} be accurately calculated in lattice QCD, and the
experimentally determined form factors. In the case of the strange
magnetic moment and charge radius of the proton, this approach has
succeeded admirably~\cite{Leinweber:2004tc,Leinweber:2006ug}. Here
we apply the technique to the strange magnetic form factor at $Q^2 =
0.23$ GeV$^2$.

Parity-violating electron scattering (PVES) has proven to be
a valuable tool
{}for experimentally determining the strange quark
contribution to the electromagnetic form factors of the proton.
Under the assumption of charge symmetry, one can deduce the strange
electric or magnetic form factor ($G_{E,M}^s(Q^2)$ ) from
measurements of the corresponding proton and neutron
electromagnetic form factors
{\it and} the neutral-weak vector form factor of the proton,
through its contribution to PVES.
While PVES measurements are
very challenging, a number of groups have succeeded, starting with
SAMPLE at Bates~\cite{Spayde:2003nr} and then A4 at
Mainz~\cite{Maas:2004dh} and
G0~\cite{Armstrong:2005hs} and
HAPPEX~\cite{Acha:2006my,Aniol:2005zg,Aniol:2004hp} at
Jefferson Lab. A global
analysis of all this data has given very precise values for the
strange quark contribution to the proton magnetic moment, as well as
its charge radius~\cite{Young:2006jc}, which are consistent with the
theoretical calculations mentioned above. The motivation for our
current work is the knowledge that in the near future we expect new
measurements from A4 and G0 at $Q^2 = 0.23$
GeV$^2$.

In addition to the extensive experimental activity, a variety of
theoretical models have been applied to the calculation of the
strange nucleon form factors. These approaches include the QCD
equalities supplemented with constituent quark model
assumptions~\cite{Derek1}, heavy baryon chiral perturbation
theory~\cite{Hemmert1,Hemmert2}, dispersive
approaches~\cite{Jaffe,Forkel,Hammer}, vector dominance model
(VDM)~\cite{Dubnicka}, VDM with a kaon cloud
contribution~\cite{Cohen}, the Skyrme model~\cite{Park}, the NJL
model~\cite{Weigel}, the chiral soliton~\cite{Silva,Goeke}, chiral
bag~\cite{Hong} and chiral quark
models~\cite{Hannelius1,Hannelius2,Lyubovitskij}, a two-component
model with a meson cloud~\cite{Bijker}, etc. These theoretical
predictions vary quite widely. For example, the predicted strange
magnetic moment varies from relatively large and negative, $-0.75\pm
0.30$~\cite{Derek1} to sizeably positive, +0.37~\cite{Hong}.

As well as the above model calculations, there have been some
lattice simulations of the strange magnetic moment, with early
lattice simulations giving a relatively large negative
value~\cite{Leinweber:1999nf,Dong,Mathur}. In 2003, Lewis $et$
$al$.~\cite{Lewis:2002ix} used low order, quenched chiral
perturbation theory, together with the lattice QCD simulation to
calculate the strange form factors from lattice data. The magnetic
form factor which they obtained at $Q^2=0.1$ GeV$^2$ was $+0.05\pm
0.06$. Recently, by combining the constraints of charge symmetry
with new chiral extrapolation techniques and low mass, quenched
lattice-QCD simulations of the individual quark contributions to the
magnetic moments of the nucleon octet, a precise, non-zero value,
$G_M^s(0)=-0.046 \pm 0.019$, was obtained~\cite{Leinweber:2004tc}.

In this paper, we present the lattice prediction for the strange
magnetic form factor at $Q^2=0.23$ GeV$^2$. We first extrapolate the
$u$ and $d$ quark contributions to the proton magnetic form factor in
quenched, heavy baryon chiral perturbation
theory~\cite{Sharpe:1992ft,Young:2002cj}.
The quenched lattice data
from the CSSM Lattice Collaboration is used and finite range
regularization (FRR) is applied in the extrapolation, because of its
improved convergence behavior at intermediate and large quark
mass~\cite{Leinweber3,Armour:2005mk,Allton:2005fb,Young2,Wang}.
In the following we briefly introduce the chiral Lagrangian which is
used in the extrapolation. The formal calculation of the magnetic
form factor is then explained, followed by the numerical results.

There are many papers which deal with heavy baryon chiral
perturbation theory. For details see, for example,
Refs.~\cite{Jenkins2,Labrenz,Durand2}. For completeness, we briefly
introduce the formalism here. In the heavy baryon chiral
perturbation theory, the lowest order chiral Lagrangian for the
baryon-meson interaction, which will be used in the calculation of
the electromagnetic magnetic form factors, including the octet and
decuplet baryons, is expressed as
\begin{eqnarray}
{\cal L}_v &=&i{\rm Tr}\bar{B}_v(v\cdot {\cal D}) B_v+2D{\rm
Tr}\bar{B}_v S_v^\mu\{A_\mu,B_v\} +2F{\rm Tr}\bar{B}_v
S_v^\mu[A_\mu,B_v]
\nonumber \\
&& -i\bar{T}_v^\mu(v\cdot {\cal D})T_{v\mu} +{\cal C}(\bar{T}_v^\mu
A_\mu B_v+\bar{B}_v A_\mu T_v^\mu) \, ,
\end{eqnarray}
where $S_\mu$ is the covariant spin-operator,
defined as
\begin{equation}
S_v^\mu=\frac i2\gamma^5\sigma^{\mu\nu}v_\nu \, .
\end{equation}
Here, $v^\nu$ is the nucleon four velocity
(in the rest frame, we have $v^\nu=(1,0)$).
$D, F$ and $\cal C$ are the coupling constants.
The chiral covariant derivative, $D_\mu$,
is written as $D_\mu B_v =
\partial_\mu B_v+[V_\mu,B_v]$.
The pseudoscalar meson octet
couples to the baryon field through the vector and axial vector combinations
\begin{equation}
V_\mu=\frac12(\zeta\partial_\mu\zeta^\dag+\zeta^\dag\partial_\mu\zeta),~~~~~~
A_\mu=\frac12(\zeta\partial_\mu\zeta^\dag-\zeta^\dag\partial_\mu\zeta),
\end{equation}
where
\begin{equation}
\zeta=e^{i\phi/f}, ~~~~~~ f=93~{\rm MeV}.
\end{equation}
The matrix of pseudoscalar fields, $\phi$,
is expressed as
\begin{eqnarray}
\phi=\frac1{\sqrt{2}}\left(
\begin{array}{lcr}
\frac1{\sqrt{2}}\pi^0+\frac1{\sqrt{6}}\eta & \pi^+ & K^+ \\
\pi^- & -\frac1{\sqrt{2}}\pi^0+\frac1{\sqrt{6}}\eta & K^0 \\
K^- & \bar{K}^0 & -\frac2{\sqrt{6}}\eta
\end{array}
\right) \, .
\end{eqnarray}
$B_v$ and $T^\mu_v$ are the velocity dependent new fields, which are
related to the original baryon octet and decuplet fields, $B$ and
$T^\mu$, by
\begin{equation}
B_v(x)=e^{im_N \not v v_\mu x^\mu} B(x),
\end{equation}
\begin{equation}
T^\mu_v(x)=e^{im_N \not v v_\mu x^\mu} T^\mu(x).
\end{equation}
In the chiral $SU(3)$ limit, the octet baryons are degenerate.
In our calculation
we use the physical mass splittings for transition meson-baryon
loop diagrams.

In the heavy baryon formalism, the propagators
of the octet or decuplet baryon, $j$,
are expressed as
\begin{equation}
\frac i {v\cdot k-\Delta+i\varepsilon} ~~{\rm and}~~ \frac
{iP^{\mu\nu}} {v\cdot k-\Delta+i\varepsilon},
\end{equation}
where $P^{\mu\nu}$ is $v^\mu v^\nu-g^{\mu\nu}-(4/3)S_v^\mu S_v^\nu$
and $\Delta=m_j-m_N$ is the mass difference between the baryon $j$
and nucleon. The propagator of meson $j$ ($j=\pi$, $K$, $\eta$) is
the usual free propagator:
\begin{equation}
\frac i {k^2-M_j^2+i\varepsilon}.
\end{equation}
\begin{center}
\begin{figure}[hbt]
\includegraphics[scale=0.4]{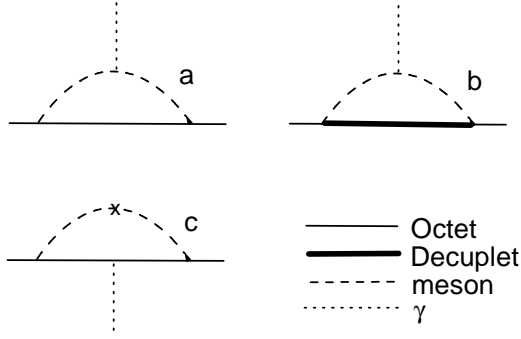}
\caption{Leading and next-to-leading order diagrams for
the proton magnetic form factors.
The last diagram, c, need only be included in the quenched
case.}
\end{figure}
\end{center}
%


In the heavy baryon formalism, the electro-magnetic form factors are
defined as:
\begin{equation}
<B(p^\prime)|J_\mu|B(p)>=\bar{u}(p^\prime)\left\{v_\mu
G_E(Q^2)+\frac{i\epsilon_{\mu\nu\alpha\beta}v^\alpha S_v^\beta
q^\nu}{m_N}G_M(Q^2)\right\}u(p),
\end{equation}
where $J_\mu$ is the charge current, $q=p^\prime-p$ and $Q^2=-q^2$.
In this paper, we focus on the magnetic form factors in each quark
sector, aiming to extract the strange quark contribution.

With the Lagrangian given earlier, the leading and next-to-leading
order diagrams for the magnetic form factor are shown in Fig.~1. In
full QCD, the first diagram, a, is the leading diagram, while
diagram b gives the next-to-leading order non-analytic term, because
of the mass difference between octet and decuplet baryons. The last,
or so-called double hair-pin, diagram need be considered only for
the quenched case, where the $\eta^{'}$ is degenerate with the pion.

The contribution to the magnetic form factor of Fig.~1a is expressed
as
\begin{equation}
\label{gea} G_M^{a}(Q^2)=\frac{{-M_N\beta}^a}{8\pi^3f_\pi^2}\int
d^3k\frac{k_y^2
u(\overrightarrow{k}+\overrightarrow{q}/2)u(\overrightarrow{k}-\overrightarrow{q}/2)}
{\omega(\overrightarrow{k}+\overrightarrow{q}/2)^2
\omega(\overrightarrow{k}-\overrightarrow{q}/2)^2}.
\end{equation}
$\omega_j(\overrightarrow{k})=\sqrt{m_j^2+ \overrightarrow{k}^2}$ is
the energy of the meson $j$. We regulate the loop integral using
finite range regularisation, with $u(\overrightarrow{k})$ the
ultra-violet regulator. Both the pion and koan are included in the
calculation. In full QCD, the coefficients are obtained from the
Lagrangian. In the quenched case the coefficients are obtained as in
Refs.~\cite{Sharpe:1992ft,Labrenz,Leinweber4}.

The contribution to the magnetic form factor of Fig.~1b can be
written as
\begin{equation}
\label{geb} G_M^{b}(Q^2)=\frac{-M_N{\beta}^b}{8\pi^3f_\pi^2}\int
d^3k\frac{k_y^2
u(\overrightarrow{k}+\overrightarrow{q}/2)u(\overrightarrow{k}-\overrightarrow{q}/2)
(\omega(\overrightarrow{k}+\omega(\overrightarrow{q}/2)+\omega(\overrightarrow{k}-\overrightarrow{q}/2))}
{A},
\end{equation}
where
\begin{equation}
A
=\omega(\overrightarrow{k}+\overrightarrow{q}/2)\omega(\overrightarrow{k}-
\overrightarrow{q}/2)
(\omega(\overrightarrow{k}+
\overrightarrow{q}/2)+\Delta)
(\omega(\overrightarrow{k}-
\overrightarrow{q}/2)+\Delta)
(\omega(\overrightarrow{k}+
\overrightarrow{q}/2)+\omega(\overrightarrow{k}-
\overrightarrow{q}/2)).
\end{equation}
In the preceding equations, $\beta^i(i=a,b)$ depends on the quark type,
meson loop type and whether the calculation involves quenched or
full QCD in the calculation.

\begin{center}
\begin{figure}[hbt]
\includegraphics[scale=0.4]{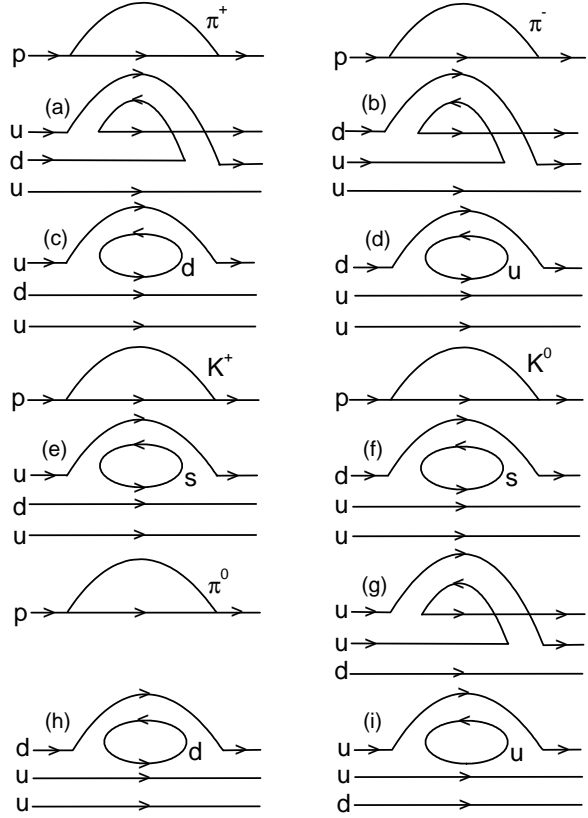}
\caption{Feynman diagrams at the quark level, which are included in
Fig.~1a for the proton magnetic form factor.}
\end{figure}
\end{center}
In the quenched case, the additional double hair-pin term from the
$\eta^\prime$ is expressed as
\begin{equation}
\label{ged}
G_M^{c}(Q^2)=\frac{(3F-D)^2M_0^2G_M(Q^2)}{288\pi^3f_\pi^2}\int d^3k
\frac{\overrightarrow{k}^2u(\overrightarrow{k})^2}
{\omega(\overrightarrow{k})^5} \, ,
\end{equation}
where $M_0$ is the double hair-pin interaction strength. We note
that the integral of Eq.~(14) gives rise to a logarithmic divergence
in the chiral limit. As a result we estimate the contribution of
this graph using the renormalized value of $G_M(Q^2)$ obtained from
the lattice simulation results at finite quark-mass values. Of
course, in full QCD no such term need be included.
\begin{table}
\caption{Coefficients, $\beta^a$, for quarks in quenched, valence
and full QCD for Fig.~1a. The left three columns are for an
intermediate $\pi$ meson and the right three columns are for an
intermediate $K$ meson.}
\begin{center}
\begin{tabular}{||c|c|c|c||c|c|c||}
quark & u & d & s & u & d & s \\ \hline Quench & $-\frac43 D^2$ &
$\frac43 D^2$
& 0 & 0 & 0 & 0 \\
Valence & $-4F^2-\frac83 D^2$ & $-\frac23 D^2+4DF-2F^2$ & 0 &
$\begin{array}{lcr} -\frac16 (3F+D)^2 ~~ \Lambda K \\
-\frac12(D-F)^2 ~~ \Sigma
K \end{array}$ & $-(D-F)^2$ & 0 \\
Full QCD & $-(D+F)^2$ & $(D+F)^2$ & 0 & $\begin{array}{lcr} -\frac16
(3F+D)^2 ~~ \Lambda K \\ -\frac12(D-F)^2 ~~ \Sigma K \end{array}$ &
$-(D-F)^2$ & $\begin{array}{lcr} \frac16(3F+D)^2 ~~ \Lambda K \\
\frac32 (D-F)^2 ~~ \Sigma K \end{array}$ \\
\end{tabular}
\end{center}
\end{table}
\begin{table}
\caption{Coefficients, $\beta^b$, for quarks in quenched, valence
and full QCD for Fig.~1b. The left three columns are for an
intermediate $\pi$ meson and the right three columns are for an
intermediate $K$ meson.}
\begin{center}
\begin{tabular}{||c|c|c|c||c|c|c||}
quark & u & d & s & u & d & s\\ \hline Quench & $-\frac{{\cal
C}^2}{6}$ & $\frac{{\cal C}^2}{6}$ & 0 & 0 & 0 &
0 \\
Valence & $-\frac{{\cal C}^2}{18}$ & $\frac{7{\cal C}^2}{18}$ & 0 &
$\frac{{\cal C}^2}{18}$ & $\frac{{\cal C}^2}{9}$ &
0 \\
Full QCD & $-\frac{2{\cal C}^2}{9}$ & $\frac{2{\cal C}^2}{9}$ & 0 &
$\frac{{\cal C}^2}{18}$ & $\frac{{\cal C}^2}{9}$ &
$-\frac{{\cal C}^2}{6}$ \\
\end{tabular}
\end{center}
\end{table}

In the above formulas, the coefficients in quenched, valence and
full QCD can be obtained with the same method as in
Ref.~\cite{Leinweber4}. For example, the diagram Fig.~1a is shown in
detail in terms of the underlying quark lines
in Fig.~2. In quenched QCD, the diagram with
a quark loop has no contribution. In the case of valence quark
sector, as well as the quenched diagram, the diagram with quark loop
can also have contribution if the external photon field couples to
the valence quark. In full QCD, both the valence and sea quark
(loop) couple to the photon field. For the pion loop, in full QCD,
Fig.~2a and Fig.~2c give contributions, while in the quenched case,
Fig.~2a and Fig.~2b give contributions. The coefficients for Fig.~2c
and Fig.~2i are the same as Fig.~2e, which is known from the
Lagrangian, since QCD is flavor blind. For the same reason, the
coefficients for Fig.~2d and Fig.2h are the same as Fig.~2f. By
subtracting the known coefficients from the total coefficients of
full QCD, we can get the coefficient for each diagram in Fig.~2.
The resulting coefficients for each quark for the different cases are
summarized in Tables I and II.

As we know, most detailed lattice simulations for the nucleon
electromagnetic form factors have been computed in the quenched
approximation, in which the strange magnetic form factor
is identically zero.
Since the value in full QCD is not large, any direct calculation of
$G_M^s$ will require considerable effort to extract an accurate
value. In this paper, we first concentrate on computing the
contribution of each valence quark to the proton form factor,
in the physical theory at the
physical mass. Then by using charge symmetry and the experimental
proton and neutron form factors, we are able to extract a precise
value of strange magnetic form factor using the techniques of
Refs.~\cite{Leinweber:1999nf,Leinweber:2004tc}.

The magnetic form factor can be expressed as
\begin{equation}
\label{ge1}
G_M(Q^2)=a_0+a_2m_\pi^2+a_4m_\pi^4+\sum_{i=a}^c G_M^{i}(Q^2) \, ,
\end{equation}
where the parameters $a_0$, $a_2$ and $a_4$ can be obtained by
fitting the quenched lattice data.
In the numerical calculations, the SU(3) parameters are chosen to be
$D=0.76$ and $F=0.50$ ($g_A=D+F=1.26$) and the coupling constant
${\cal C}$ is $ -2 D$. The FRR regulator, or form factor, $u(k)$, is
taken to be a dipole ($u(k)=\frac1{(1+k^2/\Lambda^2)^2}$, with
$\Lambda = 0.8$ GeV), although as shown by Young {\it et
al.}~\cite{Young2} the model dependence associated with other
choices is small.

We use SU(2) chiral symmetry, with only the light quark masses
varying and the strange quark
mass fixed. Thus the $K$-meson mass  is related to the pion mass by:
\begin{equation}
m_K^2=\frac12 m_\pi^2+m_K^2|_{\rm phy}-\frac12 m_\pi^2|_{\rm phy},
\end{equation}
which enables a direct relationship between
the meson dressings of
the magnetic form factor and the pion mass.

\begin{center}
\begin{figure}[hbt]
\includegraphics[scale=0.7]{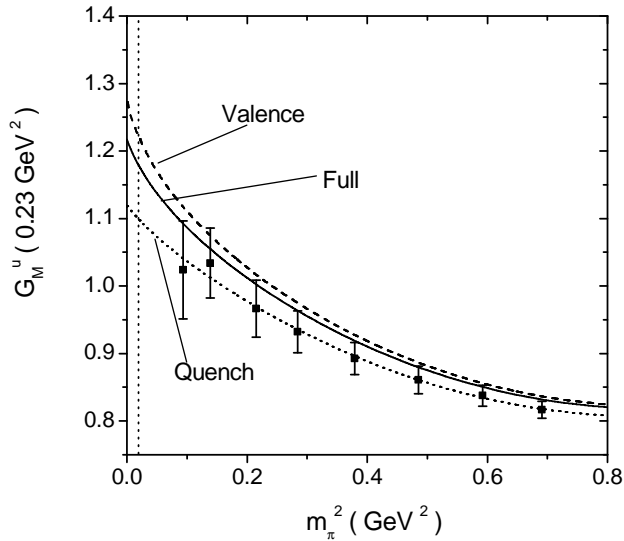}
\caption{The contribution of a single $u$ quark, with unit charge,
to the proton magnetic form factor at $Q^2=0.23$ GeV$^2$ versus pion
mass. The dotted, dashed and solid lines denote the quenched (finite
volume), valence sector and full QCD (infinite volume) results,
respectively.}
\end{figure}
\end{center}
\begin{center}
\begin{figure}[hbt]
\includegraphics[scale=0.7]{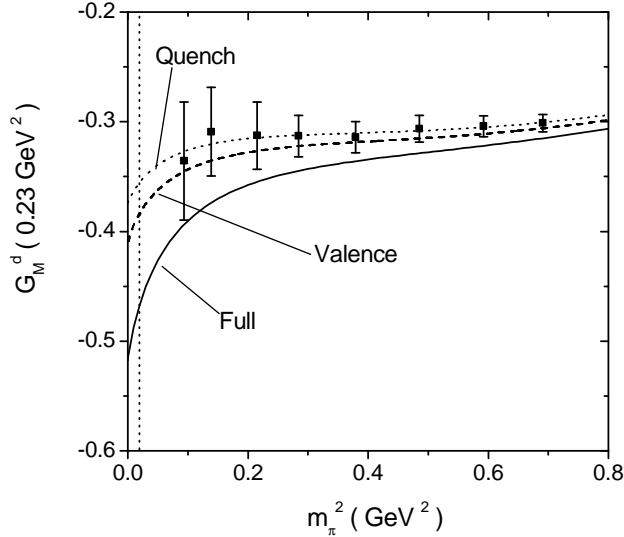}
\caption{The contribution of a $d$ quark, with unit charge, to the
proton magnetic form factor at $Q^2=0.23$ GeV$^2$ versus pion mass.
The dotted, dashed and solid lines denote the quenched (finite
volume), valence sector and full QCD (infinite volume) results,
respectively.}
\end{figure}
\end{center}
The contribution of a single $u$ quark with unit charge to the
proton magnetic form factor is shown in Fig.~3. The dotted, dashed
and solid lines are for the quenched, valence sector and full QCD
results, respectively. The square points with error bars are the
quenched lattice data obtained by the CSSM Lattice Collaboration
\cite{Boinepalli:2006xd}. The lattice results were fit with finite
volume chiral perturbation theory followed by corrections to yield
the infinite volume results. The FRR quenched chiral perturbation
theory describes the lattice data results well over the range
$m_\pi^2 \in 0.1 - 0.7$ GeV$^2$. At the physical pion mass, the
quenched ($^qG_M^u$), valence ($^vG_M^u$) and full QCD ($^fG_M^u$)
values of the magnetic form factor are $1.099 \pm 0.165$, $1.221 \pm
0.183$ and $1.179 \pm 0.177$, respectively.

In Fig.~4, we show the contribution of the $d$ quark, with unit
charge, to the proton magnetic form factor. The three styles of line
have the same meaning as in Fig.~3. Again, the quenched lattice
results are described very well. In contrast with the $u$ quark
case, the absolute value of the $d$ quark contribution in full QCD
is larger than that in the valence case. This is consistent with the
disconnected contribution and hence the strange quark form factor
being small and negative. At the physical pion mass, the quenched
($^qG_M^d$), valence ($^vG_M^d$) and full QCD ($^fG_M^d$) values of
the $d$ quark contribution are $-0.356 \pm 0.053$, $-0.383 \pm
0.057$ and $-0.468 \pm 0.070$, respectively.

With the full QCD values of the $u$ and $d$ quark contributions, one
can get the strange form factor by subtraction them from the proton
or neutron magnetic form factor. However, because of the small value
of $G_M^s$, the error bar obtained in this direct calculation is
much larger than the central value of $G_M^s$. We therefore use the
valence contributions, $^vG_M^u$ and $^vG_M^d$, which yield a
relatively precise value of $G_M^s$.

The proton and neutron magnetic form factors can be written in terms
of quark components as~\cite{Leinweber:1999nf}
\begin{equation}
\label{magp}
G_M^p = \frac43 {}^vG_M^u - \frac13 {}^vG_M^d + {}^lO_M^p,
\end{equation}
\begin{equation}
\label{magn}
G_M^n = \frac23 {}^vG_M^d - \frac23 {}^vG_M^u + {}^lO_M^n.
\end{equation}
where $^lO_M^p={}^lO_M^n=\frac23 {}^lG_M^u - \frac13 {}^lG_M^d -
\frac13 G_M^s$. The label $l$ denotes a ``loop'' or sea
quark contribution, while the label $v$ means a connected valence
quark contribution in $full$ QCD. In the equations above, charge
symmetry has been used -- i.e. the $u$ and $d$ quark contributions
in the proton are the same as the corresponding $d$ and $u$ quark
contributions in the neutron. Charge symmetry is known to be
accurate at better than 1\% where it has been tested, primarily in
nuclear systems. It has to be assumed in order to extract the
strange form factors from parity violating electron scattering.
Under the assumption of charge symmetry, the strange quark
contribution in the proton is the same as that in the neutron.

The contribution from the quark in the loop in Fig.~2 depends {\em
only} on its mass -- i.e. it is independent of whether the quark in
the loop is labelled $u, d$ or $s$. The loop contribution of each
quark can be obtained using Eqs.~(\ref{gea}) and (\ref{geb}) with
the same coefficients $\frac53D^2 - 2DF + 3F^2$ and $-\frac{{\cal
C}^2}{6}$. By calculation of the relevant loops using FRR, we
evaluate the ratio $^lR_d^s=G_M^s/^lG_M^d$ at $Q^2=0.23$ GeV$^2$.
This yields the value ${}^lR_d^s = 0.185 \pm 0.038$ allowing the
dipole mass parameter to vary between 0.6 and 1.0 GeV. Then, using
Eqs.~(\ref{magp}) and (\ref{magn}), we find
\begin{equation}
\label{gs1}
G_M^s=\frac{^lR_d^s}{1-{}^lR_d^s}(2G_M^p + G_M^n-2{}^vG_M^u),
\end{equation}
\begin{equation}\label{gs2}
G_M^s=\frac{^lR_d^s}{1-{}^lR_d^s}(G_M^p + 2G_M^n-{}^vG_M^d).
\end{equation}

In Ref.~\cite{Leinweber:2004tc}, since we were working at $Q^2 = 0$,
it was possible to use the measured magnetic moments of the nucleon
and the hyperons. Since the hyperon magnetic form factors are not
known at finite $Q^2$, here we must use the extrapolated valence
quark contributions (rather than ratios) to extract the strange form
factor. The experimental values of $G_M^p(0.23)$ and $G_M^n(0.23)$
are $\frac{G_M^p(0.23)}{\mu_pG_D(0.23)}=0.98 \pm 0.01$~\cite{Bodek}
and $\frac{G_M^n(0.23)}{\mu_nG_D(0.23)}=0.96 \pm 0.01$~\cite{JLab},
where $G_D$ is the dipole function expressed as
$G_D(Q^2)=1/(1+Q^2/0.71$GeV$^2)^2$. Substituting the experimental
magnetic moment of the proton (2.793) and neutron ($-1.913$), we
obtain the values $G_M^p(0.23) + 2G_M^n(0.23)=-0.534 \pm 0.036$ and
$2G_M^p(0.23) + G_M^n(0.23)=2.075 \pm 0.041$. Comparing the latter
with twice the value of $^vG_M^u=1.221 \pm 0.183$, obtained from our
chiral analysis of the lattice results, it is clear that there is a
significant cancellation in Eq.~(\ref{gs1}). Furthermore, the large
value of $2 ^vG_M^u$ means that the corresponding error on
$G_M^s(0.23)$ extracted from Eq.~(\ref{gs1}) will be large. Indeed,
we find that Eq.~(\ref{gs1}) yields $G_M^s(0.23)=-0.083 \pm 0.092$.
(Note that the quoted error bar arises from the errors in the
lattice data, the experimental magnetic form factors and finally the
theoretical uncertainty associated with FRR, especially the
variation of the mass parameter $\Lambda$.) On the other hand, the
relatively small value of $^vG_M^d=-0.383 \pm 0.057$ means that we
obtain a much more accurate value of $G_M^s(0.23)$ using
Eq.~(\ref{gs2}), namely $G_M^s(0.23)=-0.034 \pm 0.021$. We note that
the two extracted values of $G_M^s$ are consistent within their
respective error bars and that the sign of both, negative, is
consistent with the difference between the extrapolations of the
single quark magnetic moments in the valence and full QCD cases in
Figs.~3 and 4.

\begin{center}
\begin{figure}[hbt]
\includegraphics[scale=0.7]{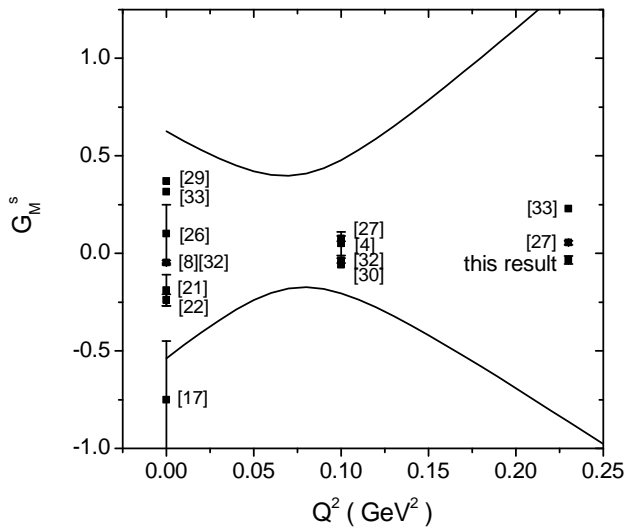}
\caption{Theoretical predictions of the strange magnetic form
factors. The two lines are for the up and low limits of the
experimental data with Eq.~(\ref{gms}).}
\end{figure}
\end{center}

Some theoretical predictions for the strange magnetic form factor
are shown in Fig.~5. These models give different values of $G_M^s$
which are all within the current experimental error bars. As for the
experimental values of $G_M^s$, using the same techniques as
Ref.~\cite{Young:2006jc}, we find:
\begin{equation}
\label{gms} G_M^s(Q^2) = 0.044 + 0.93 Q^2 \pm  \sqrt{0.34 - 7.02 Q^2
+ 47.8 Q^4},
\end{equation}
where $Q^2$ is in GeV$^2$. This form, which is the result of a
global analysis of all published data \cite{Young:2007zs}, is valid
over the range $0 < Q^2 < 0.3$ GeV$^2$.

The issue of the errors in the strange magnetic moment were, of
course, a serious issue in the earlier paper
\cite{Leinweber:2004tc}, and there and in the companion papers
\cite{Leinweber:2005bz,Thomas:2005qb} we explained all of the
sources of error, including possible charge symmetry violation. The
latter led to a much smaller contribution to the final error on
$G_M^s$ than the statistical errors on the lattice QCD data. This is
also the case here at small but finite $Q^2$. The dominance piece of
the error which we quote to $G_M^s(0.23)$ arises from the errors on
the lattice determination of $^vG_M^d$ and the experimental errors
on proton and neutron magnetic form factors, in comparison with
which the errors expected from all that is known about charge
symmetry breaking in nuclear physics, namely that it is typically
below 1\%, really are negligible\footnote{We note that the size of
the potential charge symmetry violation estimated in the calculation
of Kubis and Lewis \cite{Kubis:2006cy} is an exception, being an
order of magnitude larger than that found in the earlier calculation
by Miller \cite{Miller:1997ya,Miller:2006tv}. These authors used a
very large anomalous omega-N coupling, in contrast with what we know
from NN scattering. In addition, the omega coupling that they use
($g_\omega$) is much larger than the usual one-boson exchange
omega-N coupling. We also note that the implications of this work
for other examples of charge symmetry violation have not yet been
worked out. Nevertheless, if we were to use their extreme estimate,
our result for $G_M^s(0.23)$ would change from $-0.034 \pm 0.021$ to
$-0.025 \pm 0.024$. The difference is very small and in view of the
concerns already noted we prefer not to include this estimate of the
charge symmetry correction in our final result.}.

To conclude, we have extrapolated the lattice results for the
separate valence quark contributions to the proton magnetic form
factor at $Q^2=0.23$ GeV$^2$ in quenched and full heavy baryon
chiral perturbation theory. The leading and next-to-leading order
diagrams are considered and all octet and decuplet baryons are
included in the intermediate states. Finite-range regularisation is
used in the one loop calculation, both because it improves the
convergence of the chiral expansion and because it has been shown to
permit a connection between quenched and dynamical lattice
results~\cite{Young:2002cj}. By using the constraints of charge
symmetry, we combine the extrapolated $d$ valence quark contribution
with the experimental proton and neutron magnetic form factors to
obtain a surprisingly accurate determination of the strange magnetic
form factor $G_M^s(0.23)=-0.034 \pm 0.021$. This is the first time
it has proven possible to extract an accurate value of the strange
magnetic form factor at $Q^2=0.23$ GeV$^2$ using lattice QCD
results. It will clearly be of considerable interest to compare this
with the values which will be extracted from the recent A4 and G0
measurements at Mainz and JLab\footnote{The latest measurement of
strange quark contribution to the vector form factors was reported
after this paper was submitted for publication~\cite{A4new}. The new
result favors a negative strange magnetic form factor: $G_M^s(0.22
{\rm GeV}^2)=-0.14\pm 0.11\pm 0.11$.}.

\section*{Acknowledgements}

We thank the Australian Partnership for Advanced Computing (APAC)
and eResearch South Australia for supercomputer support enabling
this project.  This work is supported by the Australian Research
Council and by U.S. DOE Contract No. DE-AC05-06OR23177, under which
Jefferson Science Associates, LLC operates Jefferson Laboratory, and
DE-AC02-06CH11357, under which UChicago Argonne, LLC operates
Argonne National Laboratory.

\end{document}